# PERFORMANCE EVALUATION WITH A COMPARATIVE ANALYSIS OF MIMO CHANNEL ON THE BASIS OF DOPPLER SHIFT AND OTHER PROBABILISTIC PARAMETERS IN FADING ENVIRONMENT

Sutanu Ghosh

Dr. Sudhir Chandra Sur Degree Engineering College

## ABSTRACT

*At this present scenario, the demand of the system capacity is very high in wireless network. MIMO technology is used from the last decade to provide this requirement for wireless network antenna technology. MIMO channels are mostly used for advanced antenna array technology. But it is most important to control the error rate with enhanced system capacity in MIMO for present-day progressive wireless communication. This paper explores the frame error rate with respect to different path gain of MIMO channel. This work has been done in different fading scenario and produces a comparative analysis of MIMO on the basis of those fading models in various conditions. Here, it is to be considered that modulation technique as QPSK to observe these comparative evaluations for different Doppler frequencies. From the comparative analysis, minimum amount of frame error rate is viewed for Rician distribution at LOS path Doppler shift of 0 Hz. At last, this work is concluded with a comparative bit error rate study on the basis of singular parameters at different SNR levels to produce the system performance for uncoded QPSK modulation.*

## I. INTRODUCTION

Today, all the mobile users have required higher data rate with better quality of service. This higher capacity can be reached using MIMO technology [1]. MIMO is an antenna array technology with different correlation [2] pattern. Correlation is observed in between two different channels. On the basis of this correlation there are three different levels – high, medium and low. Higher amount of capacity can be achieved through low level correlation. These correlation levels are combined with different antenna array pattern – 2X2; 4X4 or, 8X8 etc. 4X4 array pattern means, each of the both end at transmitter and receiver side has four antennas. Specifically, 2X2 and 4X4 have less amount of correlation [3]. So, these two arrays are mostly used for the transmission of data. In this paper, I have used 4X4 antenna array for MIMO communication channel. Here, MIMO channel is worked with different fading model- Rayleigh fading distribution model and Rician fading distribution model. This fading is the most considerable issue for present day wireless communication system. Rayleigh [4] and Rician [5, 6] are very well known statistical distribution for amplitude modeling of radio signal in fading environment. In this research, I have worked with two different LOS path doopler shift [7] for Rician fading model. Any kind of mobile communication, antenna receives a large number of reflected and scattered waves from various directions. The instantaneous power of these received signals with Rayleigh/ Rician distribution follows the exponential function. In this work, I have considered a binary source to generate the information, which has been modulated by QPSK modulator. The information symbols output of QPSK Modulator is encoded by OSTBC Encoder [8] by using either the Alamouti code [9, 10] for two transmit antennas or other generalized complex orthogonal codes for three or four transmit antennas. The input as number of transmit antennas is given to the encoder and output of this encoder is an ($N_s$ x $N_t$) variable-size matrix, where the number of columns ($N_t$) corresponds to the number of transmit antennas and the number of rows ($N_s$) corresponds to the number of orthogonal code samples transmitted over each transmit antenna in a frame. The output of this block can be passed to the MIMO channel and finally received at OSTBC combiner [11]. The function of this combiner is to combine the received signal with different channel state information and to estimate the modulated symbols. The input signal of this combiner is an (Ns x Nr) variable-size matrix. The QPSK Demodulator block demodulates the output of OSTBC Combiner, which is a recovered modulated signal using the quaternary phase shift keying method.

Before this work, there was little research on the basis of behavior of MIMO channel. Ref. [12] introduced new algorithms for the construction of approximate minimum-error-rate linear MIMO receiver. But they didn't explain the effect of this algorithm at different level of Doppler frequency. Ref. [13] illustrated the packet error rate for different alamouti scheme [14] or receive diversity and symbol error rate for different antenna configuration with various realistic parameters. But this work didn't explain the effect of MIMO at different fading model. Ref. [15] described only the bit error rate for different level of SNR with the help of the parameters - maximum likelihood detection [16] and QPSK modulation. So, these works are not sufficient for the analysis of system performance on the problem of frame error rate at different level of Doppler frequency and different fading model. To the best of my knowledge, this kind of work has not been done for MIMO. So, I have worked on this issue to execute a performance analysis on the basis of a comparative graphical result.

The remaining portions of this paper are arranged as follows: Sections II and III, describes an overview of MIMO Channel with OSTBC Encoder-Combiner and Binary data generator with QPSK modulator-demodulator respectively. Simulation and experimental results is illustrated in Section IV. Finally, this paper is concluded in Section V.

## II. MIMO CHANNEL WITH OSTBC ENCODER AND OSTBC COMBINER:

In this work, MIMO channel is introduced with different values of Doppler frequencies. This MIMO technology have different types of antenna array configuration (2X2, 4X4, 8X8 etc) with different level of correlation as shown in Figure 1. Channel correlation is an evaluation of similarity or likeliness of two or, more different channels. There are three different correlation levels - High, Low and Medium. Here, this MIMO channel is worked with two different fading models. These models include Rayleigh and Rician distribution for different levels of Doppler frequency.

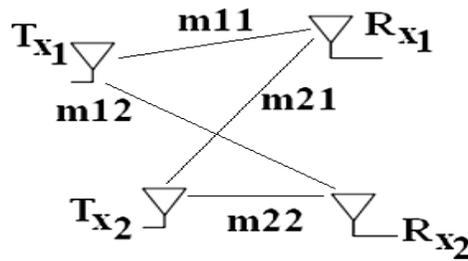

Figure 1: MIMO channel antenna configuration with different path gain m11, m12, m21 and m22

[a] Rayleigh model is mathematically expressed by Rayleigh distribution function. The probability distribution function of instantaneous power of Rayleigh model follows the exponential distribution property. The probability distribution function of power is defined by [17] –

$$p(m) = \frac{1}{m_0} \exp(-m/m_0) \qquad \ldots \ldots$$

…(1)   where, $m_0 = E[m] = \int_0^m \{m.p(m)\}dm = 2\alpha^2$

$E[m]$ is average and $2\alpha^2$ is mean square value. Rayleigh fading process can be illustrated from the mathematical manipulation over clarke's reference model [18, 19]. The low pass Rayleigh fading process [20] can be mathematically described as –

$$u(t) = \sqrt{\frac{2}{M}} \{\sum_{i=1}^{N} 2\cos(w_d t \cos\alpha_i + \Psi_i) + j\sum_{i=1}^{N} 2\cos(w_d t \sin\alpha_i + \Theta_i)\}$$

… ... (2)
where, $\alpha_i$ and $\psi_i$ are angle of incoming wave and initial phase associated with i[th] propagation path, $w_d$ is maximum angular Doppler frequency occurring when $\alpha_i = 0$. It should be characterized that selections of $\alpha_i$ and $\psi_i$ are not unique; however, different selections will proceed for different results against eq. 2.

[b] Rician, another fading model has fixed LOS component. It considers that dominant wave can be a phasor sum of two or more dominant signals (like, ground reflection, line of sight etc.). It is treated as a deterministic process. A sinusoid signal $i(t) = cos(\omega_c t)$ received over a Rician multipath channel can be characterized as –

$$o(t) = A_c \cos\omega_c t + \sum_{i=1}^{N} \alpha_i \cos(\omega_c t + \Psi_i) \qquad \ldots \ldots$$

… …(3)
where, $A_c$ is the amplitude of line of sight component (Rayleigh fading is recovered for the value of $A_c = 0$), $\alpha_i$ is the amplitude of i[th] reflected wave, $\psi_i$ is the phase of i[th] reflected wave and i =1to N identify the reflected and scattered wave. The ratio of signal power in dominant component over (local-mean) scattered power is defined as Rician $K$-factor [6].

$$K = \frac{c_m^2}{2\alpha^2} \qquad \ldots \ldots \ldots$$

…(4)
If, value of K is 0 then channel is Rayleigh and for AWGN, K is ∞.

$\alpha^2$ is local-mean scattered power and $\frac{1}{2}c_m^2$ is power of dominant component. The probability distribution function of Rician model is defined as [6] –

$$p_Z(x) = \frac{x}{\alpha^2} \exp[-(x^2 + c_m^2)/2\alpha^2] I_0(\frac{xc_m}{\alpha^2}), x \geq 0,$$

where, $2\alpha^2 = \sum_{i, i \neq 0} E[\sigma_i^2]$ is average power in non-LOS multipath components and $c_m^2 = \sigma_0^2$ is power in LOS component. The function $I_0$ is modified Bessel function of $0^{th}$ order and $Z = \sqrt{x^2 + y^2}$, where, x and y is two Gaussian random variable, both with mean 0 and equal variance.

These two fading model is associated with MIMO configuration and performed with two more blocks – OSTBC encoder [21] at transmitter and OSTBC combiner [22] at receiver side. OSTBC Encoder block encodes input symbol sequence using orthogonal space-time block code (OSTBC). Function of this block is to map input symbols block-wise and concatenates the output codeword matrices in time domain. OSTBC Encoder block supports five different OSTBC encoding algorithms on the basis of different rate and number of transmitting antennas, as shown in Table - I. There are 5 different OSTBC codeword [23] matrix for these 5 different OSTBC encoding algorithm.

Table –I : Symbol Rate Adaptation for Different Number of Transmit Antennas

| Number of transmit antennas | Symbol Rate |
| --- | --- |
| 2 | 1 |
| 3 | ½ |
| 3 | ¾ |
| 4 | ½ |
| 4 | ¾ |

This encoder supports time and spatial domains for OSTBC transmission and supports also an optional dimension; where encoding calculation is independent through that domain. This dimension may be thought of as a frequency domain.

The received information from all different receiver antennas is combined through the OSTBC combiner at receiving section. Input channel estimation may not be fixed during each codeword block transmission and combining algorithm uses only an approximation for first symbol period per codeword block. Symbol demodulator or decoder would follow Combiner block of MIMO communications system. It supports to combine each symbol independently using the combining algorithm depends on the structure of OSTBC. Combiner supports 5 different algorithm same as that of encoding algorithm. Computation algorithm per codeword block length is different for 5 different algorithms.

## III. BINARY DATA GENERATOR WITH QPSK MODULATOR AT TRANSMITTER AND DEMODULATOR AT RECEIVER OF MIMO CHANNEL

Binary data is generated by a source with Bernoulli distribution [24]. Output of this source may be frame based or sample based. Both of these two frames or sample can be expressed through a matrix. The output of this source is modulated by any kind of modulator. Here, I have used only QPSK modulator at transmitter side and demodulator at receiver side. This modulator generates a modulated symbol for every 2 successive input bits. Figure 2 depicts the total system representation of a robust MIMO channel [25 - 27] with OSTBC encoder and combiner.

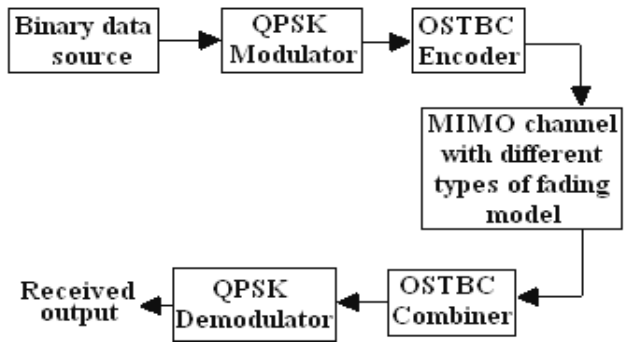

Figure 2: Block diagram of MIMO system

## IV. SIMULATION AND EXPERIMENTAL RESULTS

In this research work, I have used a set of parameters to execute the simulation. This work has been simulated in MATLAB. In this simulation work, I have considered 2 or more discrete paths to propagate signal in the system and gains of these paths are computed in decibels. Doppler velocity is utilized to measure the radial velocity of moving object. In this work, system performance is calculated on the basis of frame error rate, Doppler frequency and bit error rate with signal to noise ratio. Those parameters are as follows –

Table –II : SIMULATION PARAMETERS

| Name of the parameters | Value |
| --- | --- |
| Type of Antenna Array | 4 X 4 |
| Modulation type | QPSK |
| Type of fading model | Rayleigh and Rician |
| LOS path Doppler shift | taken as 0 and 100 Hz |

The total result can be subdivided into two different categories. One includes all the frame error rate calculations with different level of Doppler frequencies. Here, I have produced the results of frame error rate for different values of sample rate. Next section includes the bit error rate calculations for different level probabilistic parameters. This work has been simulated in MATLAB and validated through Qualnet 6.1 simulation software [28].

## RESULTS

In the first category, frame error rate has been calculated for three different cases and for all those cases, discrete path delay is assumed to be 0 value. Figure 3 shows the frame error rate for different path gain at 100 Hz Doppler shift. The graphical comparison is done on the basis of two different fading model. In these evaluations, I have considered two different conditions for Rician model on the basis of LOS path Doppler shift. It is very easy to observe the lowest frame error rate is occurred for 0 Hz Rician model.

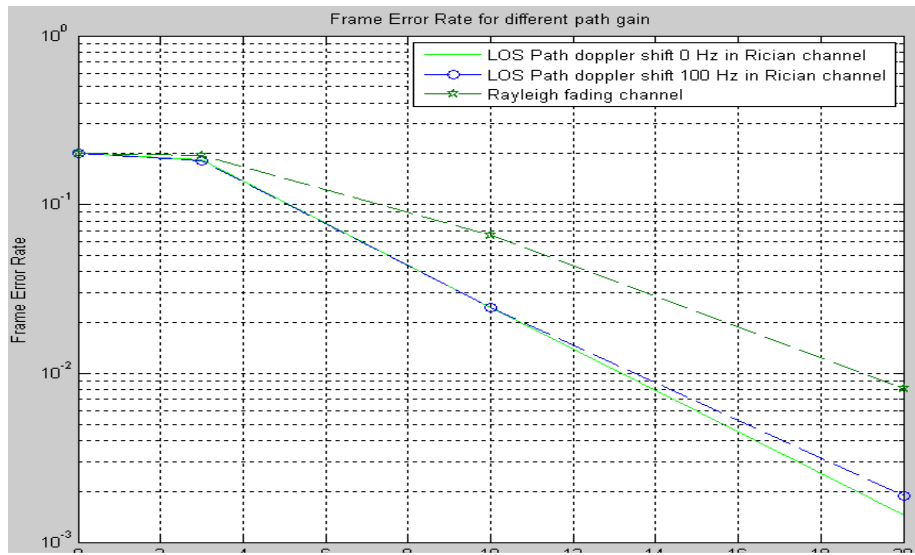

Figure 3: Graphical analysis of frame error rate for different fading model at maximum Doppler shift of 100 Hz

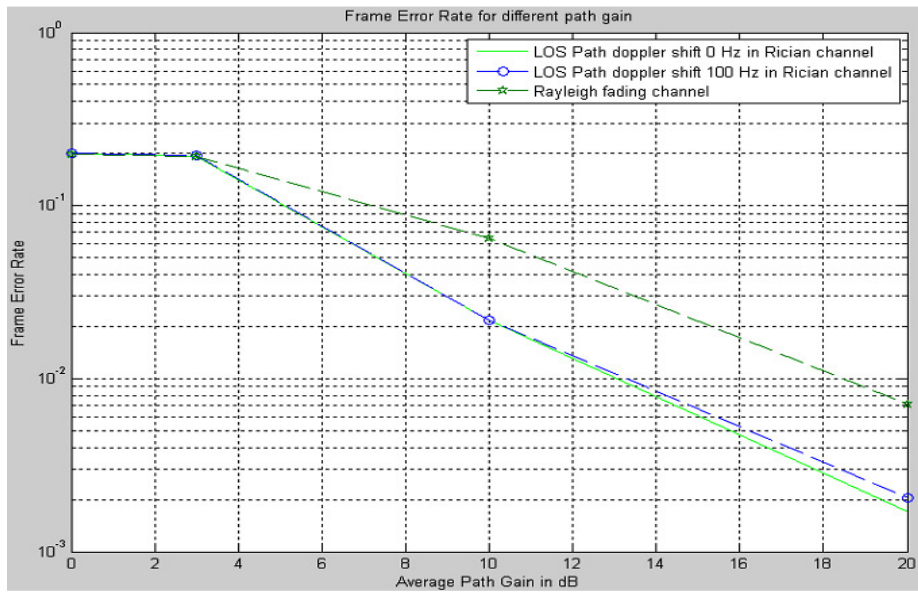

Figure 4: Graphical analysis of frame error rate for different fading model at maximum Doppler shift of 50 Hz

International Journal of Mobile Network Communications & Telematics ( IJMNCT) Vol. 4, No.5,October 2014

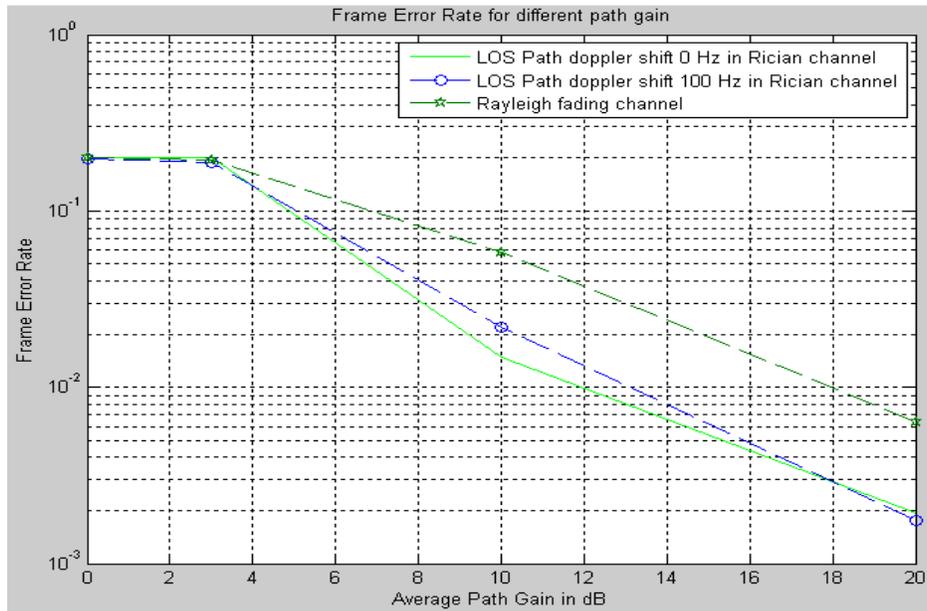

Figure 5: Graphical analysis of frame error rate for different fading model at maximum Doppler shift of 25 Hz

Figure 4 and 5 presents the comparative analysis at 50 and 25 Hz Doppler shift respectively. It is very obvious that highest frame rate is observed for Rayleigh fading model. The lowest frame error rate is viewed for LOS path Doppler shift of 100 Hz Rician fading model at 25 Hz. If the amount of Doppler shift is increased to 50 Hz and above then lowest frame error rate is observed for LOS path Doppler shift of 0 Hz Rician fading model. So, high frequency of Doppler shift is better for LOS path Doppler shift of 0 Hz Rician model on the basis of frame error rate.

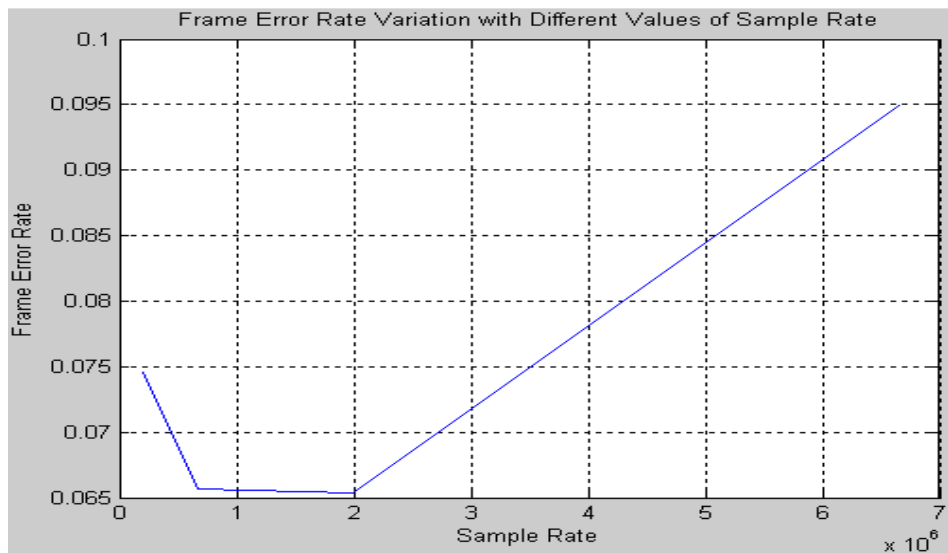

Figure 6: Graphical analysis of frame error rate for different values of sample rate in MIMO system

This Figure 6 presents the graphical result of frame error rate for different values of sample rate. This frame error rate is increased for higher values of sample rate. From initial level frame error rate is decreased upto the value of $2X10^6$, then it is increased sharply with the enhanced sample rate. The minimum value of frame error rate is observed at sample rate of $2X10^6$.

This section has described the results of bit error rate (BER) for different probabilistic variables – maximum likelihood (ML), minimum mean square error (MMSE) for different equalization scheme. ML is a mathematical algorithm to find the useful data from the noisy data stream. MMSE algorithm is used to reduce the noise power. Zero forcing (ZF) equalizer applies the inverse transfer function of the channel frequency response to received signal to bring back the signal after the channel. Here results are compared with zero forcing equalizer output and final comparative outcome is plotted in Figure 7. In this comparison, ML has the lowest BER with respect to other scheme. This ML has almost 0 BER at 20 dB SNR level.

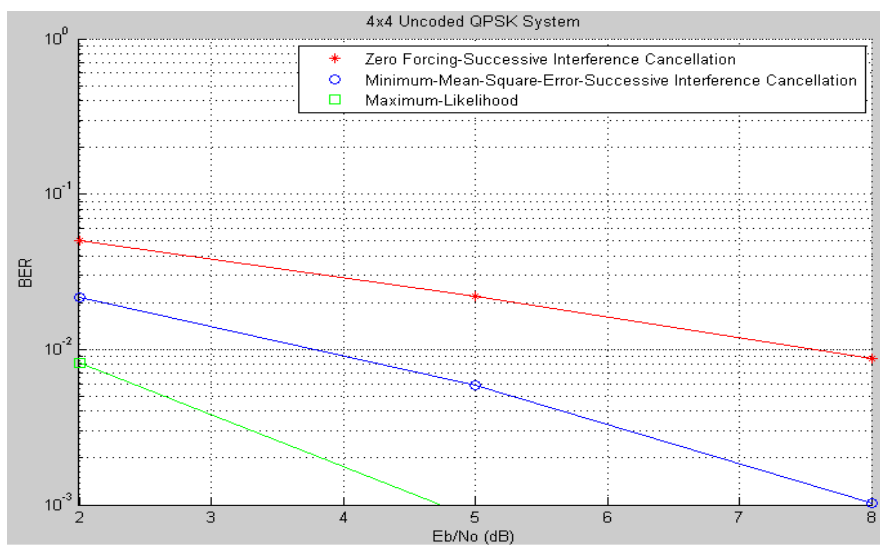

Figure 7: Graphical comparison for different pre-allocate variables in 4x4 uncoded QPSK system

## V. CONCLUSION

This paper investigates an idea about the performance evaluation on the basis of frame error rate for different frequency of Doppler shift in different types of fading model. I studied that frame error rate for Rayleigh fading is almost same for three different Doppler frequency shift (25, 50 and 100 Hz) of my experiment. From the experimental result, it can be concluded that Rician fading channel is far better than Rayleigh fading channel for the reduction of frame error rate. If the velocity of mobile object is increased at that time Rician model is performed superior than Rayleigh model.

I have produced another result for bit error rate at 4X4 system. From this result, it is proved that bit error rate is lowest and system performance is stabilized at high level of SNR for ML equalization method. The comparative result of this paper will serve a great future path for further research work.